\documentclass[aps,pre,reprint,nofootinbib]{revtex4-1}
\usepackage [latin1]{inputenc}
\usepackage{epsfig,graphics,amssymb,amsmath,subeqnarray,color}
\usepackage[sfdefault=cmbr]{isomath}
\usepackage{subfigure}
\usepackage{graphicx}
\usepackage{dcolumn}
\usepackage{bm}
\usepackage{wrapfig}
\usepackage{setspace}

\begin{document}

\title{Information transfer enhanced by non-reciprocity in a model of turning flocks }
\author{Mario Sandoval }
\email{sem@xanum.uam.mx}
\affiliation{
 Department of Physics, Complex Systems, Universidad Autonoma Metropolitana-Iztapalapa, 
 Mexico City 09340, Mexico.
}

\date{\today}
\begin{abstract}
Seminal works on animal collectives started proposing a diffusive model (overdamped) for the information transfer occurring in it \cite{Vicsek}. Afterwards, the introduction of self-rotational inertia brought into play an underdamped model able to better describe the information flux occurring in a real tuning flock event \cite{Atta}. That model was recently improved by adding nonlinear torques which allowed to match experiments \cite{cavagna2025}. The current work extends the latter model by adding active torques to a one-dimensional flock of boids (bird-like objects) while keeping key ingredients such as self-rotational inertia and nonlinearity. Those active torques are seen to enhance the system's information transfer speed and efficiency during a turning event, as well as  rendering it a non-reciprocal status. The proposed internal active torques are motivated by the adaptive injection of rotational energy (active system) of birds in a real flock  while turning. The continuum limit of the proposed model leads to a non-reciprocal modified Korteweg-de Vries (mKdV) equation with dissipation, whose structure allows the information transfer speed to be a function of the turning angular velocity. This feature occurs in real birds since under threat, birds turn faster and are required to get the information more rapid to keep cohesion. 
\end{abstract}

\maketitle

\section{INTRODUCTION}
The modeling of turning flocks is being intensively studied. Pioneering  work on this is Pomeroy and Heppner \cite{Pomeroy}  who used a three-dimensional photographie tecnique to analyze turning events in a flock of Rock doves. Consequently, one of the first active matter studies brought continuum models to describe the collective behavior of self-propelled active particles \cite{Vicsek}, as well as an attempt to understand how birds fly together \cite{Toner}. Afterwards, studies on information transfer in animal collectives \cite{Sumpter}, the governing rule of interaction among birds \cite{Ballerini}, hydrodynamic equations for interactive active particles \cite{Bertin}, a microscopic theory able to generate directional order in a flock \cite{Bia} were published. More recently, Attanasi {\it et al.} \cite{Atta} argued that information transfer in  a flock of starlings is rather underdamped than diffusive as previous models predicted. This inertial feature allows the information transfer to be less attenuated. There, they introduced the idea of self-rotation moment of inertia or the spin model \cite{Cava}. In the same year,  Yang and Marchetti \cite{HTF} used Cavagna's model \cite{Atta} 
to build its corresponding hydrodynamic equations finding spin waves (turning event in a flock) due to a coupling between self-orientation moment of inertia and elasticity. Very recently, Cavagna {\it et al.} \cite{cavagna2025} found that actually his spin model has a double peak in its  frequency domain velocity orientation correlation, which differs from their recent experimental data
showing a single peak. He then proposes the addition of a cubic torque term which stiffens the system and makes it to possess a single peak in its correlation. Other recent models obtaining a single peak in a simulated flock's velocity correlation function, have also been proposed 
\cite{liza}.
\begin{figure}
\includegraphics [width=8cm]{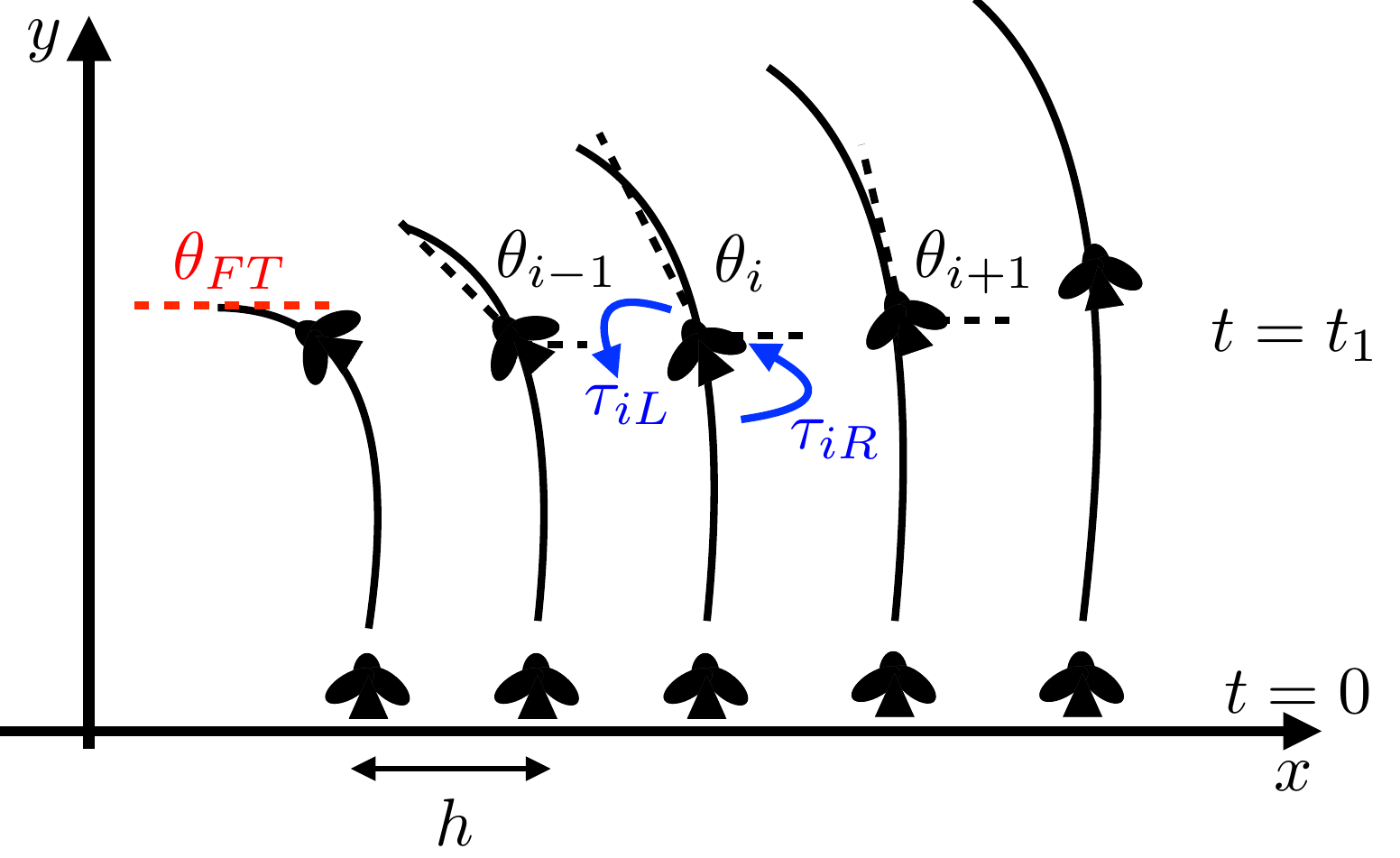}
\caption{Schematic of $N$ boids orientationally interacting by  linear and cubic nonlinear torques (characterized by constants $k$ and $\alpha$) as well as by active left $\tau_{iL}$ and right $\tau_{iR}$ torques. At $t=0$, all boids are perching along the $x-$axis. At time $t_1$ the first boid from the left is smoothly turning in the counterclockwise direction. It will reach a final turning angle $\theta_{FT}$, hence the $i-$th boid follows it with additional internal left ($\tau_{iL}$) and right ($\tau_{iR}$) turning torques which actually make the system non-reciprocal. Those internal torques are motivated by the adaptive injection of rotational energy (activity) of each bird in a real flock  while turning. }
\label{diagram}
\end{figure}
\begin{figure*}
\includegraphics [width=18cm]{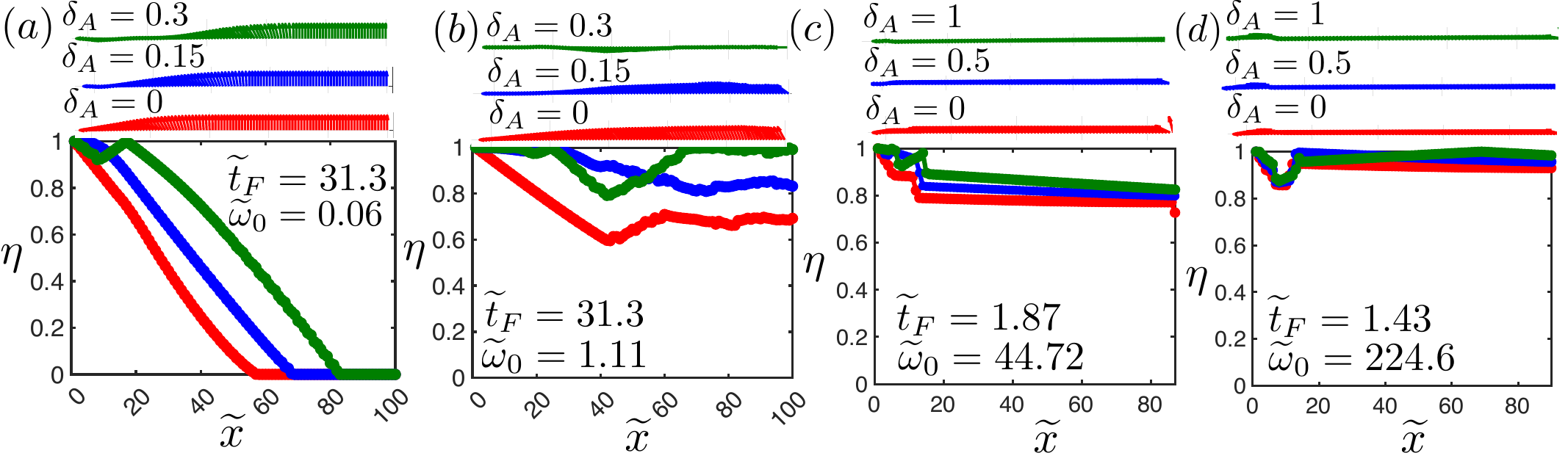}
\caption{Efficiency ($\eta$) as a function of dimensionless space ($\widetilde{x}$), non-reciprocity ($\delta_A$), and angular velocity ($\widetilde{\omega}_0$) of the initiator of the turning, for a flock of $N=100$ boids with a turning protocol from $\theta_{1}(t=0)=\pi/2$ to $\theta_{1}=\theta_{FT}=\pi$ if $t>\pi/(2\omega_0)$, and for different times. (a) Efficiency vs space for a turning dimensionless angular velocity of $\widetilde{\omega}_0=0.06$ which actually comes from experiments \cite{cavagna2025}. (b) The same as in (a) but for $\widetilde{\omega}_0=1.11$ and for different non-reciprocities. (c) The same as in (b) but for $\widetilde{\omega}_0=44.72$ and for different non-reciprocities (d) The same as in (c) but for $\widetilde{\omega}_0=224.6$. Notice that the color lines (green, blue, and red) below the $\delta_A$'s are actually showing arrows from the simulations, indicating the orientation of each boid in the flock at that time and for that non-reciprocity. }
\label{effi}
\end{figure*}
Interestingly, while performing further simulations within the 
Cavagna's nonlinear model \cite{cavagna2025}, one observes that the turning information propagation within a flock of boids (bird-like objects) \cite{Reynolds} depends on two factors: 1) the way the turning event is performed and 2) the incorporation of active torques. More explicitly, it is seen that 
the efficiency (measured here as the unity minus the error between the initial turning angle and the final boids' orientation angle) and propagation speed of a turning information event, depend on active torques and on the magnitude of the angular velocity from the boid's initiator turn. If the turning starts with a step in its information transfer (discontinuous signal), active torques effect is
negligible. On the other hand, if the turning starts with a smooth information signal (akin to real turning flocks without any threat \cite{cavagna2025}), the signal propagation efficiency is improved by the incorporation of active torques, which in fact, render the system a 
non-reciprocal status. Notice that the novelty in the current work is the proposal of active torques, which is motivated by the fact that birds during a turning event, add both translational and rotational energy (aerodynamic forces) \cite{rotation} at the individual level to the flock.
It is also worth mentioning that a spin model with added non-reciprocity in the way of  a step function has been reported very recently \cite{Bandini}. The current work can also be seen as a 1D spin model made non-reciprocal due to adaptive active torques.
\section{TURNING FLOCK MODEL}
\label{model}
\begin{figure*}
\includegraphics [width=18cm]{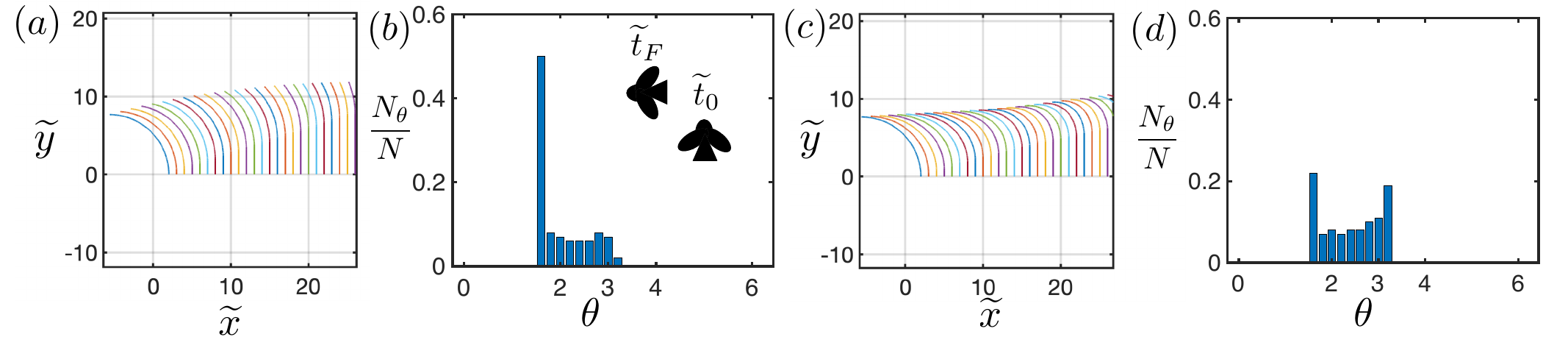}
\caption{Turning of $N=100$ boids --orientationally interacting by  linear and cubic nonlinear torques-- from $\theta_{i}(\widetilde{t}=0)=\pi/2$ to $\theta_{1}=\theta_{FT}=\pi$, as indicated in the inset. (a),(c) Paths of the turning birds from $\widetilde{t}=\widetilde{t}_0=0$ until $\widetilde{t}=\widetilde{t}_F=31.3$ with $\delta _{A}=\{0,0.2\}$, respectively. (b) Paths' dispersion angle at $\widetilde{t}=31.3$ with  $\delta _{A}=0$. (d) Paths' dispersion angle at $\widetilde{t}=31.3$ with  $\delta _{A}=0.2$. Clearly, non-reciprocity makes boids to align faster with the turning signal $\theta_{1}=\theta_{FT}=\pi=3.1415$ as the histograms show.  }
\label{angle}
\end{figure*}

Consider  $N$ boids (bird-like-objects) initially perching along the $x-$axis and separated by a distance $h$. Let each $i-$th boid dynamics be given by its orientation angle $\theta_i$ and location $\mathbf{x}_i=(x_i,y_i)$. Without loss of generality (otherwise an interacting adjacency matrix should be included), assume that they orientationally interact only between its 
nearest neighbors following a linear + cubic nonlinear torque model, see \cite{FPUO,ZABUSKY1967}, and more recently, Cavagna \cite{cavagna2025}. However, given the fact that real birds are active in both translational and orientational degrees of freedom \cite{rotation}, it is proposed that Cavagna's idea may become orientationally active if one adds internal active torques, $\tau^{int}$, exerted by particle $i$. Let us also propose that the strength of
the added internal active torques linearly depend on the  instantaneous net angular difference measured by 
the $i-$th boid from its left and right hand side (this idea can actually be thought of as the addition of  adaptive active torques to the model \cite{rotation}), that is, $\tau_{iL}=-\tau^{A}(\theta_{i}-\theta_{i-1})$ and 
$\tau_{iR}=-\tau^{A}(\theta_{i+1}-\theta_{i})$, respectively, with $\tau^{A}$ a constant. The negative sign is to have active torques mainly acting towards the {\bf k} direction, since it will be proposed that the turning event has this direction. The opposite direction in the turning will simply need a change of sign in the active torques. For a clearer picture, see
Fig. \ref{diagram}. Note as well that translational self-propulsion by each boid is incorporated by the speed $v_0$ along each boid's orientation unit vector $\mathbf{e}_i(t)$. Thus it is assumed that the translational dynamics of the $i-$th boid satisfies 
\begin{equation}
\frac{d\mathbf{x}_i}{dt}=v_0\mathbf{e}_i(t),  \label{trans} 
\end{equation}%
where $\mathbf{e}_i(t)=(cos(\theta_i),\sin(\theta_i))$, while the rotational degree of freedom ($\theta_i$) satisfies
\begin{align}
\chi\frac{d^2\theta_i}{dt^2}=& -\gamma\frac{d\theta_i}{dt}  \notag
\\
& -k(\theta_{i}-\theta_{i-1})-\tau^{A}(\theta_{i}-\theta_{i-1})  \notag
\\
& +k(\theta_{i+1}-\theta_{i})-\tau^{A}(\theta_{i+1}-\theta_{i})  \notag \\
& -\alpha (\theta_{i}-\theta_{i-1})^{3}+\alpha (\theta_{i+1}-\theta_{i})^{3}.  \label{ori}
\end{align}%
where $k$ and $\alpha $ define respectively, the linear and nonlinear
part of the interacting torque, $\chi$ represents the spin (self-rotation) moment of inertia (different from the inertial angular moment), while  $\gamma$ indicates friction. Note the physical  interpretation of the
latter equation. It adds the idea that  whenever a boid senses an instantaneous angular difference, it  will manoeuvre in the  {\bf k} angular direction proportionally to that difference, both towards the turning signal direction and away from its right-hand side neighbor.
The latter leaded to postulate  the left internal torque ($\tau_{iL}$) and the right internal one ($\tau_{iR}$), both mainly acting along the {\bf k} direction in accordance to the turning signal. This is further clarified in Fig. \ref{diagram}, where one can see the pair of internal active torques (blue  curved arrows) generated by the $i-$th boid and pointing counterclockwise. The magnitude of those left and right active torques may also be weighted; however for simplicity, we will keep them the same in this model.
Furthermore, let us assume that the torque coefficient
satisfies, $\tau^{A}=k\delta _{A}$, with $\delta _{A}$  a constant, thus Eq. (\ref{ori}) can be written as
\begin{align}
\chi\frac{d^2\theta_i}{dt^2}=& -\gamma\frac{d\theta_i}{dt}  \notag
\\
& -k(1+\delta _{A})(\theta_{i}-\theta_{i-1})+k(1-\delta
_{A})(\theta_{i+1}-\theta_{i})  \notag \\
& -\alpha (\theta_{i}-\theta_{i-1})^{3}+\alpha (\theta_{i+1}-\theta_{i})^{3}.
\label{springs}
\end{align}%
The previous dynamics can be interpreted as the breakdown of Newton's third law. Notice that  $-k(1+\delta _{A})(\theta_{i}-\theta_{i-1})$
is the reaction torque from boid $i-1$ to $i$; however, the action torque from boid $i$ to $i-1$ according to this model is, $k(1-\delta _{A})(\theta_{i}-\theta_{i-1})$. Clearly Newton's third law is violated. This violation then
leads to a non-reciprocal response \cite{Martin,Coulais2} of  the current system.  In summary,  our current model is out-of equilibrium in both translational and rotational dynamics, and non-reciprocal due to the incorporation of active torques. The following sections will show the effect of the addition of active torques by presenting the system's information transfer efficiency, dispersion angle, and information transfer speed. %
\section{EFFICIENCY OF THE INFORMATION TRANSFER}
\label{effisec}
Let us define the efficiency $\eta_i$ of the information transfer for each $i-$th boid as
\begin{equation}
\eta_i=1-\varepsilon_i,
\end{equation}
where $\varepsilon_i \in [0,1]$ stands for the error typically defined as $\varepsilon_i=(\theta^*_{FT}-\theta^*_{iF})/\theta^*_{FT}$. Here, $\theta^*_{FT}=\theta_{FT}-\theta_{I}$ represents the signal  propagation  final angle, $\theta^*_{iF}=\theta_{FT}-\theta_{I}$ indicates the final angle of a given boid, while $\theta_{I}$ is the initial orientation given to all boids. In this way, the information transfer efficiency can now be calculated for each boid. The system considered here has the following numerical parameters 
$N=100$, $\chi=1$Js, $k=20$J/s, $\alpha=100,000$J/s, $\gamma=0.7$J, $\omega_0=\{0.26,5,200,1000\}$s$^{-1}$, $dt=0.001$s, and $\delta_A=[0,1]$ which are practically the same as in \cite{cavagna2025} and were selected in such a way to mimic the dynamics of a real turning flock. If those parameters are not properly selected, additional oscillations in the system due to an excess of  rotational energy may be observed. For generalization purposes nondimensional quantities will be introduced, namely, $\widetilde{x}=x/h$, $\widetilde{t}=t\sqrt{k/\chi}$, $\widetilde{\omega}_0=\omega_0\sqrt{\chi/ k}$, and $\widetilde{c}_0=c\sqrt{\chi/ k}/h$. Note that other flocks with $N=\{50,100,200\}$ where also studied and the same results were obtained, thus proving that the current system is size independent. To start the simulations, all the $i$-th boids initiated with $\theta_{i}(t=0)=\theta_{I}=\pi/2$. The turning protocol was selected as a uniform angular displacement only for the first boid in the row, that is, $\theta_{1}=\omega_0 t + \pi/2$ if $t<\pi/(2\omega_0)$, otherwise, $\theta_{1}=\theta_{FT}=\pi$.  The rest of the boids will follow that boid initiator. This protocol will constantly be shown as an inset in the figures. The numerical procedure follows a vectorized second order Verlet algorithm. The results are shown in Fig. \ref{effi} in  a dimensionless way. Figures \ref{effi}(a), (b), (c) and (d) illustrate the flock efficiency at different times, $\widetilde{t}_F=\{31.3,3.57,3.26\}$, respectively, and for different non-reciprocity values ($\delta_A$). Note that  $\widetilde{t}_F=31.3$ was chosen since it is the time a real flock without threat takes to make a U-turn \cite{cavagna2025}. The other times were taken since at those instants, the flock of boids is practically aligned to the left. Colors above the plots are actually arrows (directly from simulations) indicating the state of the system, respectively. These figures in general indicate that at small  turning angular velocities, non-reciprocity or active torques have a clear effect, that is,  the higher the active torque, the better the efficiency in the system will be. However, those plots also indicate that, active torques will play a small role for  higher turning angular velocities, since the efficiency in the system stays around 90$\%$ and practically independent from $\delta_A$, see for example Fig. \ref{effi}(d). It is worth mentioning that Fig. \ref{effi}(a) is the simulation of a 1D (one-dimensional) flock with parameters roughly taken from experiments (see Cavagna's video \cite{cavagna2025} and also Section \ref{dispe}). This figure clearly shows an efficiency improvement due to non-reciprocity. See how at $\delta_A=0.3$, the turning information has reached the $N=35$ boid with $90\%$ of accuracy, whereas for the same boid but with $\delta_A=0$, the efficiecny has dropped to  $40\%$, and hence boid $N=35$ is still half the way, namely, $\theta_{35}=3\pi/4$ (see Fig. \ref{effi}(a) red arrows below $\delta_A=0$). Additionally, one can also see from Fig. \ref{effi}(b)-(d) that the current model predicts that boids under threat will eventually increase their angular velocity, which in turn will enhance its information transfer efficiency. This feature is positive in a real flock since it ensures cohesion and thus a higher probability of survival.
 \begin{figure}
\begin{center}
\includegraphics [width=8cm]{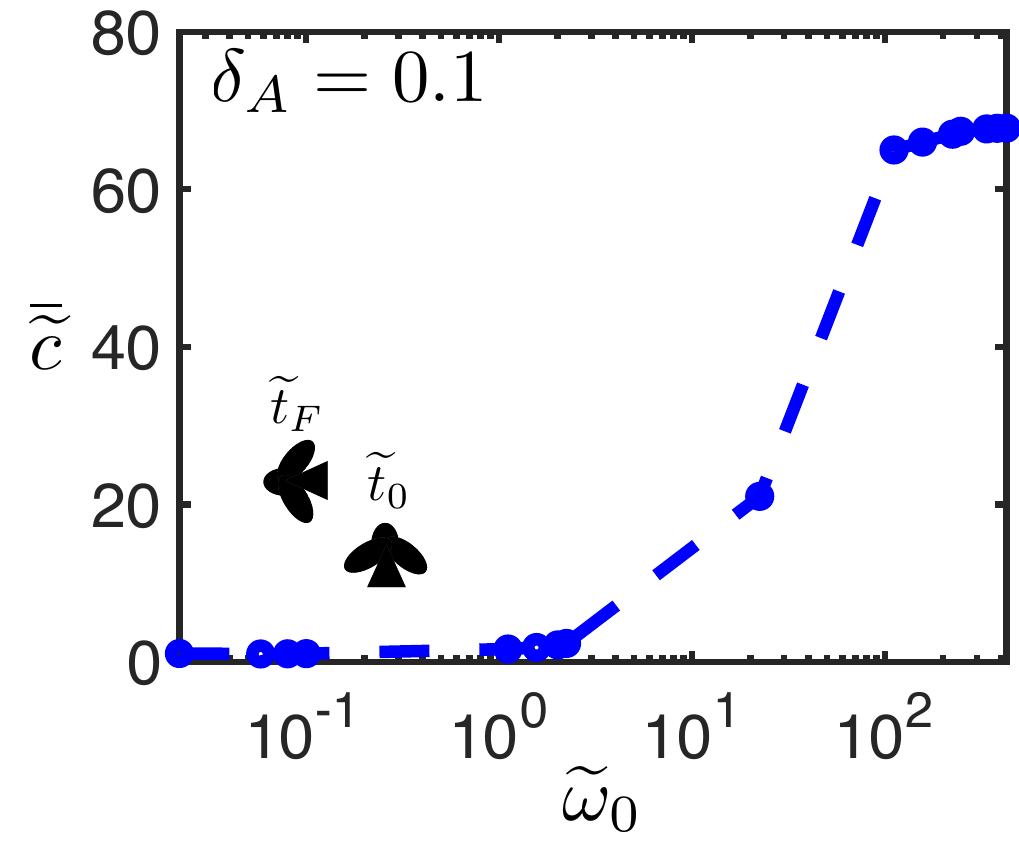}
\caption{Average dimensionless propagation signal speed vs dimensionless angular velocity. Here, the non-reciprocal constant is taken as $\delta_A=0.1$. Other $\delta_A$ value gives practically the same behavior. The turn protocol is from $\theta_{0}=\pi/2$ to $\theta_{FT}=\pi$, as indicated in the inset. Notice that for small and high angular velocities, the propagation signal remains practically constant.}
\label{speed}
\end{center}
\end{figure}
\section{DISPERSION ANGLE MEASUREMENT}
\label{dispe}
From Cavagna's video \cite{cavagna2025}, it is possible to infer that the turning time of a real flock whithout any threat last around $7$s. From that video, one can also roughly extract the flock's angular velocity, which is around $\omega_0=0.26$rad/s. Thus in dimensionless form, that flock's turn is approximately characterized by  $\widetilde{\omega}_0=0.06$ and $\widetilde{t}_F=31.3$. Those parameters will be used to simulate under the current model,  the turning of  a flock of boids. The results at $\widetilde{t}_F=31.3$ are illustrated in Fig. \ref{angle}. Looking at Fig. \ref{angle}(a) and Fig. \ref{angle}(c), where typical boid's paths are drawn, one can infer that the lack of active torques (Fig. \ref{angle}(a)) makes the transfer of information less efficient, since the orientation angles of each boid are more spread with respect to the turning event final angle, $\theta_{FT}=\pi$. This is further indicated by the histogram in Fig. \ref{angle}(b). Here, the orientation angles are away from $\theta_{FT}=\pi$. On the other hand, Fig. \ref{angle}(c) shows that the boids' paths are closer to  $\theta_{FT}=\pi$ (around $N=30$ boids are completely facing  to the left), hence one can see that the incorporation of active torques ($\delta_A=0.2$) improves the transfer of 
information. This is also illustrated in the histogram from Fig. \ref{angle}(d), where now there are more boids at an angle $\theta_{FT}=\pi$. This last behavior is closer to what one observes in nature, where birds quickly turn as long as a turning signal has started. Notice as well in Fig. \ref{angle}(a) and Fig. \ref{angle}(c), the semicircle paths formed by the boid's initiator turn, which are due to the implemented turning protocole.
\section{PROPAGATION SIGNAL speed AS A FUNCTION OF NON-RECIPROCITY}
\begin{figure}
\includegraphics [width=8cm]{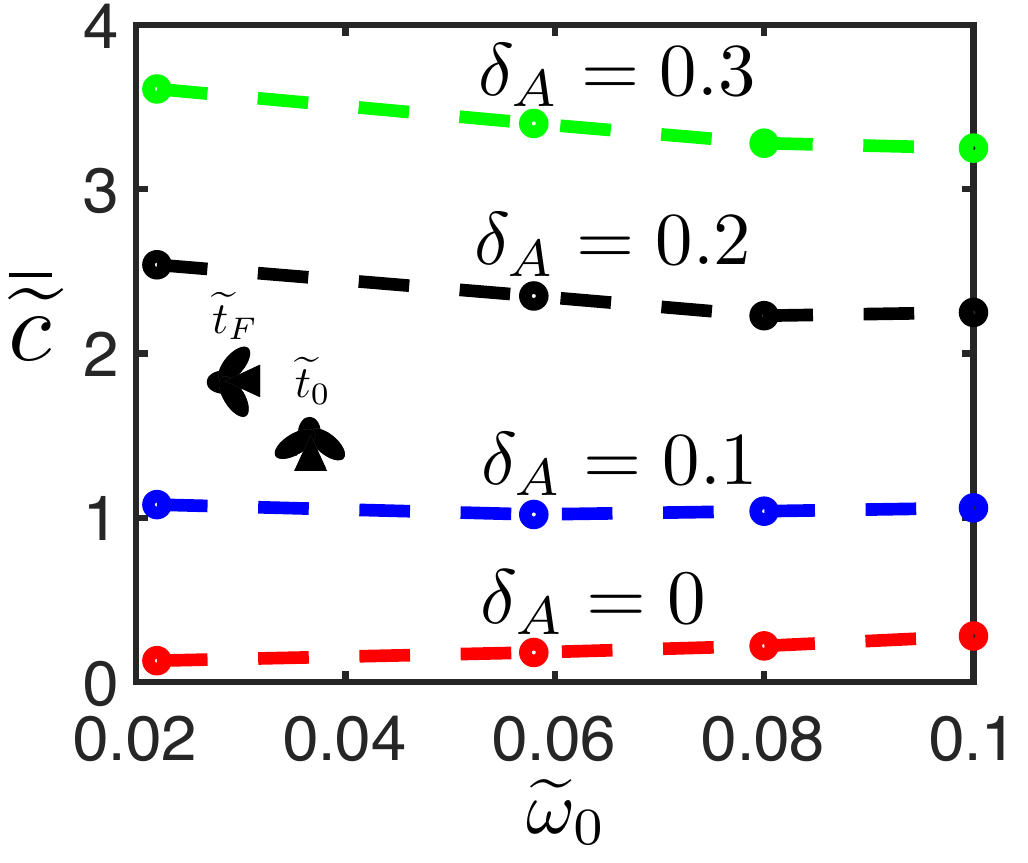}
\caption{Average dimensionless propagation signal speed vs dimensionless angular velocity and for different non-reciprocal constants, namely, $\delta_A=\{0,0.1,0.2,0.3\}$. The turn protocol is the same as in previous figures. For this studied case, notice how non-reciprocity is able to increase the propagation signal by at least an order of magnitude.}
\label{delta}
\end{figure}

Let us now study if non-reciprocity due to the presence of active torques modify the propagation signal speed. To do so, different turning angular velocities were considered as indicated in Fig. \ref{speed} and Fig. \ref{delta}, and the location in time  of the propagation turn throughout the boids' flock was tracked.  Basically, the idea is to count the time a given $i-$th boid reaches within a $2\%$ error the $\theta_{FT}$ angle, and then to plot position of the propagation signal vs time. It was observed that at small angular velocities ($\widetilde{\omega}_0\leq10^1$) and small active torques, the motion of the propagation signal may be accelerated; and that for larger angular velocities ($\widetilde{\omega}_0>10^1$) and irrespective of $\delta_A$, its motion becomes linear. For that reason, one introduces the average dimensionless propagation speed 
$\overline{\widetilde{c}}$, which is an average of the instantaneous velocities the propagation signal has, in a position vs time plot. This nonlinear propagation is due to the proposed turning protocol (slow rotation) and to the existence of dissipation, otherwise a linear signal propagation should be observed. The result is illustrated in Fig. \ref{speed}, where
$\overline{\widetilde{c}}$ vs $\widetilde{\omega}_0$ and for $\delta_A=0.1$ (other $\delta_A$ will practically have the same behavior apart from the region of small $\widetilde{\omega}_0$) is plotted. One can see that for small angular velocities ($\widetilde{\omega}_0\leq10^1$), the average propagation signal is around the unity and constant. The same happens at  $\widetilde{\omega}_0>10^1$ but now, there seems to be a boundary layer between those regions. After many simulations, one notices that in order to see an effect of active torques on the propagation speed, one has to analyze the region $\widetilde{\omega}_0<10^1$. Larger angular velocities showed a small effect. The result is illustrated in Fig. \ref{delta} where $\overline{\widetilde{c}}$ vs $\widetilde{\omega}_0$ and for $\delta_A=\{0,0.1,0.2,0.3\}$ is offered. Clearly, at these small angular velocities, active torques have indeed an effect on the propagation signal  information. This figure indicates that the larger the non-reciprocity, the faster the signal. It actually grows an order of magnitude since for $\delta_A=0$ one has $\overline{\widetilde{c}}\sim0.1$, whereas for $\delta_A=0.3$ we have $\overline{\widetilde{c}}\sim3.2$. Interestingly, the experimental angular velocity ($\widetilde{\omega}_0=0.057$) from Cavagna's video falls within the domain of Fig. \ref{delta}, hence one can see why under these parameters, non-reciprocity affects the turning event of our simulated flock of boids (compare Fig. \ref{angle}(a) and Fig. \ref{angle}(c) for example).
In order to give a mathematical explanation of the results in Fig. \ref{speed} and Fig. \ref{delta}, let us take the continuum limit of Eq. (\ref{springs}). By Taylor expanding it,  one gets the following non-reciprocal
nonlinear wave equation 
\begin{equation}
\frac{1}{c_{0}^{2}}\theta_{tt}+\overline{\gamma} \theta_{t}=(1+\epsilon \theta_{x}^{2})\theta_{xx}-\frac{%
2\delta _{A}}{h}\theta_{x}+\frac{h^{2}}{12}\theta_{xxxx},
\end{equation}%
where  $\overline{\gamma}=\gamma/h^2k$ , $c_{0}=\sqrt{kh^2/\chi}$ and $\epsilon=3h^2\alpha/k$. The latter expression can be further transformed under the change of variables $\xi =x-c_{0}t$, 
$\tau =\epsilon c_{0}t/2$, together with $\theta_{\xi }=\psi,\eta =\theta_{\tau }$, which to
$O(\epsilon )$ leads to
\begin{align}
\psi _{\tau }-\frac{\overline{\gamma} c_{0}}{2}\eta +\psi ^{2}\psi _{\xi }+& \frac{h^{2}%
}{12\epsilon }\psi _{\xi \xi \xi }+\left( \frac{\overline{\gamma} c_{0}}{\epsilon }-%
\frac{2\delta _{A}}{h\epsilon }\right) \psi =0,  \label{d1} \\
& \quad \psi _{\tau }=\eta _{\xi }.  \label{d2}
\end{align}%
which is a coupled system in $\psi(\xi,\tau)$ and $\eta(\xi,\tau)$ due to the presence of
dissipation, and with the structure of  a non-reciprocal modified Korteweg-de Vries (mKdV) equation \cite{Sandoval}. That PDE (partial differential equation) is well-known to accept solitons whose speed depend on the initial amplitude, which is precisely what we observe in Fig. \ref{speed} since in our case, the angular velocity plays the role of amplitude. Equation (\ref{d1}) also tell us that there are terms accounting for energy injection ($\frac{2\delta _{A}}{h\epsilon }\psi $) and dissipation ($  \frac{\gamma c_{0}}{\epsilon }\psi $) \cite{Bona}, which are candidates to possibly change the propagation of information. This change was confirmed in Fig. \ref{speed}  for $\widetilde{\omega}_0\leq10^{-1}$ where the injection of energy accounted by $\delta_A$ proportionally increases the information transfer propagation signal. As it can be seen, the structure of system (\ref{d1})-(\ref{d2}) confirms our numerical findings, and whose  behavior, suggest that the propagation signal in  a turning event is actually  a non-reciprocal soliton. Moreover, one can also get some physics after an order of magnitude analysis of the injection and dissipation energy terms is performed. Based on dimensional values (\cite{cavagna2025} and Section \ref{effisec}), one gets for $\delta_A=0.3$, that the term $\frac{2\delta _{A}}{h\epsilon }=4\times10^{-4}m^{-3}$, whereas the term $  \frac{\overline{\gamma} c_{0}}{\epsilon }=1.04\times10^{-4}m^{-3}$, thus implying that the system is receiving energy (a signature of active matter). Due to this very small difference between activity and dissipation, the  resulting soliton information signal may accelerate. Note that the effect of 
$\omega_0$ will appear in the initial condition needed when solving the system (\ref{d1})-(\ref{d2}). 

\section{Summary and conclusions}
\label{summa}
It has been shown that active torques indeed increase the signal propagation speed when an initial turning event has angular velocities in the order of $\widetilde{\omega}_0<10^1$, which is the order of magnitude of real birds turning without any threat (0.26rad/s or in dimensionless form $\widetilde{\omega}_0=0.057$) \cite{cavagna2025}. 
Then, active rotation --leading to non-reciprocity-- may also be an ingredient for explaining why a flock seems to immediately turn when one member has already started. It was also seen that due to the nonlinearity in the present model, a high turning angular velocity (which may exemplify a boid under threat) is able to increase the system's propagation signal for several orders of magnitude, which is a response needed in a real flock to keep cohesion within the group. It was also observed that the propagation speed signal for turning events with a high angular velocity ($\widetilde{\omega}_0>10^2$), is practically unaffected by active torques since its efficiency for even $\delta_A=0$ is already above $80\%$. In perspective, this paper contributes to the modeling of information transfer in animal collectives that started as an overdamped diffusive model \cite{Vicsek}, which was consequently improved with the idea of self-rotational inertia (spins) making it underdamped \cite{Atta}, then by incorporating nonlinear torques to match experiments \cite{cavagna2025}, and currently, by adding active torques seen to enhance the information  transfer speed and efficiency.

\section{Acknowledgements}

M. S. thanks the Mexican working community whose taxes
support this research through Secretaria de Ciencia, Humanidades, Tecnologia
e Innovacion (Secihti).

\bibliography{flub}

\end{document}